# Neuroplasticity in adult human visual cortex

Elisa Castaldi[1], Claudia Lunghi[2] and Maria Concetta Morrone[3,4]

[1] Department of Neuroscience, Psychology, Pharmacology and Child health, University of Florence, Florence, Italy

[2] Laboratoire des systèmes perceptifs, Département d'études cognitives, École normale supérieure, PSL University, CNRS, 75005 Paris, France

[3] Department of Translational Research and New technologies in Medicine and Surgery, University of Pisa, Pisa, Italy

[4] IRCCS Stella Maris, Calambrone (Pisa), Italy

# Abstract

Between 1-5:100 people worldwide has never experienced normotypic vision due to a condition called amblyopia, and about 1:4000 suffer from inherited retinal dystrophies that progressively lead them to blindness. While a wide range of technologies and therapies are being developed to restore vision, a fundamental question still remains unanswered: would the adult visual brain retain a sufficient plastic potential to learn how to 'see' after a prolonged period of abnormal visual experience? In this review we summarize studies showing that the visual brain of sighted adults retains a type of developmental plasticity, called homeostatic plasticity, and this property has been recently exploited successfully for adult amblyopia recover. Next, we discuss how the brain circuits reorganizes when visual stimulation is partially restored by means of a 'bionic eye' in late blinds with Retinitis Pigmentosa. The primary visual cortex in these patients slowly became activated by the artificial visual stimulation, indicating that sight restoration therapies can rely on a considerable degree of spared plasticity in adulthood.

# Highlights

- ➤ Short-term monocular deprivation triggers homeostatic plasticity in adult humans
- ➤ Physical activity and inverse occlusion promote visual recovery in amblyopic adults
- ➤ Intra-modal and cross-modal brain reorganization occur in late-blind individuals
- ➤ Sight restoration in late-blinds promotes early visual cortex re-organization
- ➤ The adult visual system retains residual plasticity well after the critical period

# Index

# 1. Introduction

Neuroplasticity is the ability of the nervous system to adapt and optimize its limited resources in response to physiological changes, injuries, new environmental demands and sensory experiences (Pascual-Leone et al., 2005).

Early in the individual's development, during the so called *critical period*, the intrinsic plastic potential of the nervous systems is maximal and sensory deprivation events can cause profound morphological alterations of sensory cortices preventing the normal development of sensory functions (Berardi et al., 2000; Wiesel and Hubel, 1963). Classical studies in the visual system of kittens and non-human primates demonstrated that even few days of abnormal visual experience during the period of high susceptibility can cause severe visual impairment that cannot be recovered even after prolonged periods of restored vision (Hubel and Wiesel, 1970; Hubel et al., 1977; Wiesel and Hubel, 1963, 1965b). Visual cortical plasticity is typically assessed by using the experimental paradigm of *monocular deprivation*: occluding one eye few weeks after birth causes atrophy of the lateral geniculate nucleus (LGN) layers representing the deprived eye, reorganization of primary visual cortex (V1) ocular dominance columns, with ocular dominance shifting in favor of the open eye and with only a few neurons responding to the deprived eye (Hubel and Wiesel, 1970; Hubel et al., 1977; Wiesel and Hubel, 1963). At the functional level, this shift of ocular dominance is reflected in lower visual acuity and reduced response to stimulation of the deprived eye, a condition known as *amblyopia* (for a review see: Levi and Carkeet, 1993). In adults, after the closure of the critical period, amblyopia cannot be induced or treated: in adult cats monocular deprivation has only minor or no effects, suggesting that the visual cortex retains very little or no experience-dependent plasticity in adulthood (Wiesel and Hubel, 1963). Similarly, in humans early visual deprivation or suboptimal visual experience (for example due to untreated congenital cataracts, astigmatism, myopia, etc.) leads to plastic reorganization of the visual cortices and permanent visual impairments that can hardly be recovered in adulthood even after corrections or visual restoration (Braddick and Atkinson, 2011; Fine et al., 2002, 2003; Maurer et al., 2005, 2007). The ocular dominance imbalance induced by early altered monocular inputs can be efficiently treated by occluding the dominant eye for prolonged periods of time (occlusion therapy) but only within the critical period (Webber and Wood, 2005), whereas only modest improvements of visual function are observed when therapy is performed in adulthood (Fronius et al., 2014).

While neuroplasticity is preserved in adulthood for higher-level functions, endorsing learning and memory (Fuchs and Flugge, 2014), the absence of experience-dependent changes observed after the closure of the critical period led researchers to consider the visual system, and in particular early visual cortex, as hard-wired and with no spared plastic potential, especially for ocular dominance. However, several recent lines of evidence have put this assumption into question: in animal models, ocular dominance plasticity can be reactivated after the closure of the critical period by manipulating the visual cortex excitability, either pharmacologically or through environmental enrichment and physical activity (Baroncelli et al., 2010, 2012; Berardi et al., 2015; Harauzov et al., 2010; He et al., 2006; Hensch and Fagiolini, 2005; Hensch et al., 1998; Maya Vetencourt et al., 2008; Pizzorusso et al., 2002; Sale et al., 2007, 2014; Spolidoro et al., 2009). In adult humans, evidence of preserved visual plasticity has been demonstrated by behavioral and neural changes associated with perceptual learning (Beyeler et al., 2017; Dosher and Lu, 2017; Fiorentini and Berardi, 1980; Watanabe and Sasaki, 2015), short term visual deprivation (Binda et al., 2018; Binda and Lunghi, 2017; Lunghi et al., 2011, 2013, 2015a,b, 2018; Lunghi and Sale, 2015; Zhou et al., 2013, 2014), progressive blinding pathologies and visual restoration therapies (Aguirre et al., 2016; Baseler et al., 2011a; Burton, 2003; Castaldi et al., 2016; Cunningham et al., 2015a,b; Dormal et al., 2015; Heimler et al., 2014).

The cellular mechanisms promoting neuroplasticity during development have been extensively studied in animals (Smith et al., 2009). Cortical wiring in developing neural systems is determined both by molecular cues, responsible for neural migration and formation of synaptic contacts, as well as by activity-dependent mechanisms that fine tune and optimize the number and strength of synaptic connections through *Hebbian plasticity* (Levelt and Hubener, 2012). This is mediated by Long Term Potentiation and Depression (LTP/LTD), mechanisms that have a crucial role also in learning and memory (Malenka and Bear, 2004) and in defining the critical period (Berardi et al., 2000; Hensch and Quinlan, 2018; Levelt and Hubener, 2012). In addition Hebbian plasticity, another form of plasticity (named *homeostatic plasticity*) is acting to maintain the overall balance of the network excitability (Maffei and Turrigiano, 2008a). This may have an important role given that LTP/LTD themselves might lead to destabilization of neural circuits: continuous feedback cycles might lead to the progressive strengthening/weakening of synaptic connections and to its consequent excessive excitability/loss (Turrigiano and Nelson, 2000). Through mechanisms such as synaptic scaling, synaptic redistribution and changes in neural excitability, the homeostatic plasticity promotes stability in the neural circuits by readjusting the overall level of network activity to optimize responses to sensory experiences and perturbations (Abbott and Nelson, 2000; Maffei and Turrigiano, 2008a,b; Turrigiano, 2012; Turrigiano and Nelson, 2000). After short-term deprivation and during the critical period homeostatic plasticity is boosting the signals from the deprived eye, in the attempt to contrast the deprivation effects. Recent evidence suggests that this homeostatic synaptic rescaling may remain active even after the closure of the critical period and support neuroplasticity throughout the life span. Interestingly, the visual cortex maintains homeostasis by normalizing the individual neural responses with the overall activity of a pool of neurons. This mechanisms is especially important to efficiently control contrast gain (Carandini and Heeger, 2011; Heeger, 1992). It is possible that homeostatic plasticity in adult humans may be implemented using similar cellular mechanisms mediating contrast gain (see later Lunghi et al., 2011)

Here, we review evidence for adult plasticity and we speculate on the most likely mechanisms and principles supporting the residual ocular dominance plasticity in adulthood. Deeply understanding the mechanisms regulating adult visual plasticity is crucial for visual restoration in late-blind individuals, considering that the various intervention (retinal prostheses but also pharmacological manipulations) are implemented monocularly: in principle the restored monocular signals might be gated at cortical level if ocular dominance plasticity cannot be endorsed.

We start reviewing recent studies reporting perceptual and neural changes induced in healthy sighted individuals by temporary altering their visual experience and how these findings have recently led to the development of a non-invasive training for adult amblyopic patients. We discuss how these phenomena likely reflect homeostatic plasticity in adulthood. In the second part we discuss the neural changes following blinding diseases and in particular the plastic retinotopic remapping of the residual visual input and the cross-modal reorganization of the visual brain when the visual input is altered or interrupted. Finally, we describe the success and falls of the very first attempts of restoring vision in adult blind patients.

## 2. Cortical plasticity in sighted adults revealed by short-term visual deprivation

### 2.1 Behavioral proxies for plasticity

Recent studies have introduced new behavioral techniques to infer and estimate the degree of residual plasticity for ocular dominance in adult sighted subjects, by combining binocular rivalry (Lunghi et al., 2011) and, more recently, pupillometry (Binda and Lunghi, 2017) with short periods of monocular deprivation. *Binocular rivalry* is a form of bistable perception that is generated whenever two incompatible images are separately projected to the eyes. In such condition the visual perception alternates between the two monocular images which take turn in dominating visual awareness (Blake and Logothetis, 2002; Levelt, 1965). Binocular rivalry is one of the most robust psychophysical methods used to assess sensory eye dominance (Ooi and He, 2001): under normal conditions, the average time in which the image presented to each eye dominates the observer's perception is similar for the two eyes, reflecting balanced ocular dominance. Lunghi et al. (2011) first observed that visually depriving healthy adult individuals though monocular occlusion with a translucent patch for 150 minutes profoundly altered ocular dominance measured with binocular rivalry. Surprisingly the stimulus presented to deprived eye dominates twice as long as the one displayed in the non-deprived eye after deprivation. The effect, although progressively attenuated, lasted up to 90 or 180 minutes after patch removal (Lunghi et al., 2013). After patch removal the apparent contrast increased for a short time, suggesting an up-regulation of the contrast gain-control mechanisms in the occluded eye that boosted the neuronal responses to compensate for the reduced incoming signal. The altered dynamics of binocular rivalry and the contrast gain enhancement after monocular deprivation suggested the existence of a spared plastic potential in adult individuals: these phenomena most likely reflect a form of homeostatic plasticity attempting to maintain the overall network activity stable and to optimize the individual's new visual experience.

Even though the perceptual effects of short-term monocular deprivation might be in principle interpreted as contrast adaptation, growing evidence indicates that they reflect a genuine form of plasticity. For example, even though deprivation alters apparent contrast, the 36% boost in apparent contrast is not sufficient to explain the change in ocular dominance (Lunghi et al., 2011), and when chromatic vision is specifically targeted (using iso-luminant visual stimuli), the effect can out-last the duration of deprivation, lasting for up to 3h (Lunghi et al., 2013). Moreover, two recent studies (Bai et al., 2017; Ramamurthy and Blaser, 2018) have shown that changes in ocular dominance can be observed by altering the monocular input without reducing monocular contrast, pointing to a crucial role of inter-ocular correlation in mediating ocular dominance plasticity. This is consistent with the etiology of amblyopia which can be produced by strabismus (Kiorpes et al., 1998) where both eye receive good vision but with no spatial congruency. Another direct proof that the transient changes in ocular dominance are an expression of homeostatic plasticity is given by the long term improvement of vision in adult amblyopic patients that received a short-term monocular deprivation of the amblyopic eye for 6 short sessions. The improvement in visual acuity of about 2 lines lasted for up to one year, indicating long-term neuroplastic changes (Lunghi et al., 2018).

Binda & Lunghi (2017) more recently tackled another biomarker of neuroplasticity in adult humans by demonstrating that monocular deprivation affects spontaneous low frequency oscillations of the pupil diameter at rest, a phenomenon called *hippus* (Diamond, 2001). The authors measured pupillary oscillations before and after monocular occlusion, following the same paradigm as Lunghi et al. (2011), and found that hippus amplitude increased after visual deprivation and that participants with more pronounced pupillary fluctuations also showed stronger ocular dominance changes in binocular rivalry dynamics. The procedures to measure these two proxies of neuroplasticity are completely non-invasive and suitable for application in clinical populations.

## 2.2 Short-term visual deprivation alters visual neural responses

A substantial number of studies investigating the cortical effects of visual deprivation in adult humans suggested that plasticity might be mediated by changes in the excitation/inhibition balance of the visual cortex (Binda et al., 2018; Boroojerdi et al., 2000; Fierro et al., 2005; Lou et al., 2011; Lunghi et al., 2015a; Pitskel et al., 2007).

The first studies (Boroojerdi et al., 2000; Fierro et al., 2005; Pitskel et al., 2007) applied TMS pulses to the occipital cortex sufficient to elicit light perception (phosphene) in absence of visual stimulation. The minimum intensity needed to elicit phosphene perception (PT, phosphene threshold) is an indirect measure of cortical excitability. Boroojeredi et al. (2000) showed that after 45 minutes of binocular light deprivation PT was reduced in healthy adult subjects, suggesting increased cortical excitability and this effect persisted for the entire deprivation period (180 minutes). Re-exposure to light reverted the process and PT returned to pre-deprivation values within 120 minutes. Interestingly the same study provided also another measure of cortical responsiveness, independent of the subjects' perception: neural activity in striate and extrastriate cortices as measured by BOLD responses with functional magnetic imaging (fMRI) was enhanced after 60 minutes of blindfolding and the increased fMRI signal persisted for at least 30 minutes after re-exposure to light.

Monocular deprivation might cause different cortical effects with respect to blindfolding, possibly due to inter-ocular competition which is absent in case of binocular deprivation. In fact, evidence from animal studies has shown that, compared with monocular deprivation, binocular deprivation performed during the critical period induces a more modest reorganization of visual cortical circuits (Wiesel and Hubel, 1965a). In adult humans, monocular deprivation was shown to cause a decrease in cortical excitability, rather than an increase, as measured by TMS PT (Lou et al., 2011). However a recent study measuring pattern-onset visual evoked potentials (VEPs) before and after monocular deprivation described a more complex pattern of results with opposite effects for the two eyes (Lunghi et al., 2015a). The amplitude of the C1 component of the VEP responses, typically reflecting the earliest stage of visual processing in V1 (Di Russo et al., 2002), and the peak in alpha band after monocular deprivation were enhanced for the deprived eye, but reduced for the non-deprived eye. Moreover the amplitude of later components, such as P1 and P2, were equally altered by monocular deprivation, suggesting that the variations in cortical excitability propagate to extrastriate areas and modulate feed-back projection to V1. Overall these results, in line with the perceptual changes observed after short-term monocular deprivation (described in section 2.1), suggest that the deprived and non-deprived eye are respectively strengthened or weakened after monocular deprivation, reflecting antagonistic homeostatic short-term plasticity in the two eyes.

Recently Binda et al. (2018) exploited the enhanced resolution and signal to noise ratio provided by ultra-high field (7T) fMRI, to track ocular driven changes of BOLD responses in V1 before and after 2h of monocular deprivation. During the scanning participants' eyes were separately stimulated with either high contrast low- and high-band-pass noise (optimized to differentially stimulate the magno- and parvocellular pathways respectively) or with a luminance matched uniform background (Fig.1A). BOLD responses to the high spatial frequency stimuli were strongly affected by monocular deprivation, in opposite directions for the two eyes: the percentage BOLD signal measured in V1 elicited by stimulation of the deprived eye and non-deprived eye was respectively enhanced and reduced with respect to pre-deprivation values (Fig 1B). The authors further calculated for each voxel an index of Ocular Dominance defined as the response difference to the deprived and non-deprived eye, which reflects the eye preference of a given particular voxel (average biased signal), although not strictly coinciding with the ocular dominance columns. Before deprivation, the ocular dominance indices were symmetrically distributed around zero, reflecting balanced V1 activity elicited by the two eyes. However, after deprivation the index strongly shifted toward a stronger activation from the deprived eye. Interestingly the voxels originally preferring the deprived eye did not change their preference (i.e. the average signal measured in these voxels continued to be biased toward the same eye), whereas the voxels originally preferring the non-deprived eye swapped their preference and were most strongly activated by stimulation of the deprived eye. Importantly the deprivation effect in BOLD responses correlated with the perceptual effect of deprivation as measured by increased mean phase duration for the deprived eye during binocular rivalry performed outside the scanner with the classic paradigm (Fig. 1 C, D). Spatial frequency selectivity in V1 was also affected by deprivation: only responses elicited by high-spatial frequencies stimulation of the deprived eye were significantly reduced, while responses to low-spatial frequencies stimulation of the same eye as well as to the whole frequencies range of the non-deprived eye were unaffected. Population Spatial Frequency Tuning of V1 confirmed these effects and the shift of selectivity toward higher spatial frequency in the deprived eye correlated with the phase duration during binocular rivalry. Interestingly, the increase in BOLD response to the high-spatial frequency stimulation of the deprived eye was strongest in V1, V2, V3, attenuated but still significant in V4, while it was absent in V3a and hMT+. The enhanced signals measured along the ventral but not along the dorsal pathway suggested that plasticity acted more strongly on the parvocellular rather than magnocellular pathway. This interpretation is also supported by the evidence that deprivation-induced changes in binocular rivalry dynamics with chromatic equi-luminant gratings resulted in much longer-lasting effects with respect to those measured with luminance-modulated gratings (Lunghi et al., 2013).

In sum, these experiments showed that complex patterns of neuroplastic responses can be induced in the visual cortex of sighted adults by short periods of visual deprivation, which rapidly alter the visual cortex's excitability and responsiveness to homeostatically adapt to the altered visual experience.

## 2.3 Neurochemical changes following short-term visual deprivation

The molecular mechanisms underlying experience-dependent plasticity have been widely investigated in animals (Berardi et al., 2003; Heimel et al., 2011). The maturation of intracortical inhibition has been proved to play a crucial role in regulating the progression of critical period for ocular dominance and visual acuity (Berardi et al., 2003; Fagiolini et al., 2004; Fagiolini and Hensch, 2000; Hensch et al., 1998; Huang et al., 1999; Speed et al., 1991). For instance, increasing intracortical inhibition was shown to anticipate the opening and closure of the critical period for monocular deprivation in mice (Fagiolini and Hensch, 2000; Hanover et al., 1999; Huang et al., 1999), and transgenic animals lacking a GABA-synthesizing enzyme showed deficient ocular dominance plasticity to monocular deprivation which could be restored by increasing inhibitory transmission with benzodiazepines (Hensch et al., 1998). In adulthood, the increased inhibition may be a limiting factor for cortical plasticity and reducing GABAergic inhibition was shown to partially restore ocular dominance plasticity in adult rats and promote recovery from amblyopia (Harauzov et al., 2010; Maya Vetencourt et al., 2008). Interestingly, environmental enrichment and physical activity were recently found to be associated with reduced inhibitory tone in the rat's visual cortex, providing a potential non-invasive strategy to promote recovery from amblyopia (Baroncelli et al., 2010, 2012; Sale et al., 2007; Stryker, 2014).

In line with these results, one study recently showed that GABAergic inhibition plays a key role in promoting ocular dominance plasticity also in adult humans (Lunghi et al., 2015b). Participants underwent one psychophysical test measuring ocular dominance by means of binocular rivalry and one 7T MR spectroscopy session before and after 150 min of monocular occlusion (same procedure as described in section 2.1). The perceptual changes triggered by deprivation (resulting in the deprived eye dominating over the non-deprived eye) were associated with decreased GABA concentration in V1, at least when GABA concentration was assessed at rest, while participants kept their eyes closed. Specifically, participants with greater decrease in resting GABA concentration showed the greatest perceptual effects, i.e. perceptual boost and dominance of the deprived eye. Interestingly, other studies showed that fMRI responses in visual cortex inversely correlated with GABA concentration, potentially suggesting a link between GABA level and cortical excitation-inhibition balance (Donahue et al., 2010; Muthukumaraswamy et al., 2009).

Of course GABAergic circuits are not the only ones mediating plasticity in visual cortex (Berardi et al., 2003). In animals, ocular dominance plasticity can be restored also by enhancing excitatory neurotransmission systems such as serotoninergic (Maya Vetencourt et al., 2008) and cholinergic (Morishita et al., 2010) systems. In humans

Boroojeredi et al. (2001) applied TMS over occipital cortex and measured phosphene thresholds in adult sighted participants under the effect of various drugs interfering with synaptic plasticity before and after a period of light deprivation. They found that the rapid plastic changes typically triggered by light deprivation (here quantified as phosphene detection thresholds) were blocked when participants were under the effect of lorazepam (which enhance the functioning of GABAa receptors), dextromethorphan (NMDA receptors antagonist) and scopolamine (a muscarinic receptor antagonist), thus pointing at a key role of GABA, NMDA and cholinergic receptors in mediating rapid plastic changes after visual deprivation.

Finally, in light of the recent results by Binda & Lunghi (2017) showing both increased pupillary hippus at rest and enhanced eye-dominance of the deprived eye during binocular rivalry after monocular deprivation, the role of the neurotransmitter norepinephrine (NE) in mediating homeostatic plasticity should be further investigated in adult humans. This neurotransmitter may indeed constitute the common source of visual cortical excitability underlying these phenomena, given its known role in regulating both pupil diameter modulation (Joshi et al., 2016) and visual cortical plasticity (Kasamatsu et al., 1979, 1981).

Overall, these studies point at a spared plastic potential in the adult visual cortex beyond the critical period which can be reactivated by altering the excitability level of visual cortex either by pharmacologically targeting several neuromodulator systems or by manipulating sensory and motor experience, such as being exposed to abnormal visual experiences even for a short time period, or being exposed to an enriched environment and physical activity.

## 2.4 Therapy for amblyopia

Inspired by studies on animal models showing that physical exercise triggers visual cortical plasticity, modulates visual cortex excitability and increases neurotrophic factors (Baroncelli et al., 2010, 2012; Sale et al., 2007), recent studies tested whether similar effects could be obtained in adult humans (Lunghi and Sale, 2015; Lunghi et al., 2018). Lunghi & Sale (2015) tested binocular rivalry dynamics in sighted participants before and after monocular deprivation (still with the same paradigm described in 2.1) while varying the level of physical activity performed by participants during the deprivation period. In the 'inactive condition' participants were required to watch a movie while sitting on a chair, while in the 'physical activity' condition they watched a movie while intermittently cycling on a stationary bike. With respect to the control inactive condition, the perceptual effect of deprivation on binocular rivalry dynamic was much stronger when participant performed physical activity throughout the 2h period tested after eye-patch removal, suggesting that physical activity had further boosted homeostatic plasticity.

The beneficial effect of moderate physical activity for triggering neuroplasticity was recently combined with short-term inverse occlusion to promote the visual recovery in adult anisometropic amblyopes (Lunghi et al., 2018). Six 2h long training sessions over a 4 weeks period consisting in simultaneous physical activity (intermitted cycling) and occlusion of the amblyopic eye restored visual acuity in all patients and stereopsis in six of them with improvement lasting up to 1 year. Although these results should be replicated in a larger sample, this non-invasive training paradigm seemed to successfully boost visual plasticity in adulthood and to constitute a valid approach to treat amblyopia.

# 3. Cortical plasticity after vision loss following ophthalmological diseases

## 3.1 Retinotopic remapping of the visual cortex

The blind brain certainly constitute a unique opportunity for studying plasticity in adulthood. A highly studied phenomena, sometimes considered as an index of neuroplasticity, is the ability of the visual cortex to remap the retinotopic organization of the neuronal receptive fields following retinal lesions (for a recent review see: Dumoulin and Knapen, 2018). The first evidence for this phenomenon was reported in adult cats and monkeys (Gilbert and Wiesel, 1992): after retinal lesion V1 receptive fields near the border of the lesion projection zone (LPZ) underwent an immediate enlargement, and two months after the regions silenced by the lesion were found to represent loci surrounding the scotoma. The receptive fields shift leading to complete filling in of the scotoma was attributed to long-range horizontal connections within V1 rather than to spared geniculate afferent connections: none of the changes observed in the recovered cortex where observed in the lateral geniculate nucleus, perhaps due to its reduced plasticity with respect to higher level cortical regions (however see section 3.3).

More recently these findings have been strongly questioned by a combined neurophysiological and fMRI study that failed to record normal responsivity in adult macaque V1 during 7.5 months of follow-up after retinal lesion (Smirnakis et al., 2005). The authors found no change in the BOLD-defined LPZ border and suggested that cortical reorganization is not needed to explain the apparent size of the LPZ (Smirnakis et al., 2005). Incongruences with respect to previous results were attributed to several factors, including sampling biases and differences in the recording methods (one recording sub-sets of single neurons, the other reflecting average activation of ensembles of cells) and originated an intense debate (for details on this debeat see: Calford et al., 2005; Sereno, 2005; Smirnakis et al., 2005; Wandell and Smirnakis, 2009).

The ability of the adult visual cortex to reorganize has also been studied in retinal dystrophies, namely macular degeneration (MD, both age-related MD and juvenile MD) and retinitis pigmentosa (RP). Both diseases create scotomas in patients' central (MD) and peripheral (RP) visual field that expand with illness progression. Some fMRI studies described large-scale cortical reorganization of visual processing in response to retinal disease by showing that the regions of V1 matching the patients' scotoma are remapped to respond to stimuli outside it both in MD (Baker et al., 2005, 2008; Dilks et al., 2009, 2014; Schumacher et al., 2008) and in RP patients (Ferreira et al.,

2017). However, others have found the LPZ to remain silent (Ritter et al., 2018; Sunness et al., 2004) and found no evidence for large scale remapping of the visual cortex following late blindness (Baseler et al., 2011b; Goesaert et al., 2014; Haak et al., 2015). Masuda et al. (2008; 2010) found that the possibility to detect BOLD signals in LPZ depends on task and might reflect the upregulation or unmasking of feed-back projections from extrastriate areas to V1 that are normally suppressed in sighted individuals. These authors proposed that the presence of activations in LPZ might not need to reflect cortical reorganization. Similar conclusions were reached by studies that simulated scotomas in sighted humans artificially by removing the visual stimulus in a given location of a rich image. This studies found consistently altered receptive fields properties around the simulated scotoma (Binda et al., 2013; Dumoulin and Knapen, 2018; Haak et al., 2012).

In sum, studies in patients with retinal dystrophies have reported mixed results regarding the presence of remapped visual activation in LPZ. Identifying the origin of such activations remains matter of open debate: it can reflect cortical reorganization leading to the formation of new connections or changes in the connections strength between neurons or to unmasking of existing connections normally suppressed in sighted individuals.

### 3.2 Cross-modal plasticity in blind individuals

Another widely studied phenomena, considered a marker of experience-dependent neuroplasticity, is the cross-modal reorganization of the visual cortex in blind individuals: several studies observed occipital activations elicited by non-visual sensory stimulation in congenital and early-onset blinds (Amedi et al., 2010; Burton, 2003; Collignon et al., 2009, 2011b; Ptito et al., 2005). The mechanisms leading such cross-modal reorganization are still under debate. Some authors proposed that cross-modal activations in blinds reflect unmasking of 'latent' cross-modal connections that are normally suppressed (Merabet et al., 2007, 2008; Pascual-Leone and Hamilton, 2001), while others pointed at an additive shifts of the cross-modal responses in early-blind individuals rather than at rescaling or unmasking processes (Fine and Park, 2018; Lewis et al., 2010). Interestingly the cross-modal reorganization of typically visual areas is not random, but seems to rather reflect a supramodal functional organization of the brain, encoding for an abstract representation of the perceived stimuli independently from the sensory input (Pietrini et al., 2004; Ricciardi et al., 2014a,b; Ricciardi and Pietrini, 2011). For example, responses in the typically visual motion area MT+ can be elicited by motion-specific auditory and tactile stimulation in early-blinds and sight recovered individuals (Ricciardi et al., 2007; Saenz et al., 2008). However the overlap between cross-modal responses should be taken carefully and further investigated to exclude artefactual co-localizations of cross-modal responses within the same area (Jiang et al., 2015). Yet, signals reflecting functional selective cross-modal plasticity have been reported by several studies across different cortical areas and cognitive functions (such as object recognition, stimuli localization in the space, reading) and seems to be a general reorganizational principle of the brain independently from the deprived modality (Heimler et al., 2014): areas typically involved in spoken language processing for example, are recruited by sign language in deaf people (MacSweeney et al., 2008).

The presence of cross-modal responses in late-onsets blind individuals is more controversial. For instance, Bedny et al. (2010) found activations elicited by auditory motion stimuli in hMT+ only in congenital, but not in late-blind subjects. Similarly the perceived direction of auditory moving stimuli could be classified in early, but not in late-blind individuals (Jiang et al., 2014, 2016) using multi voxel pattern analysis. However, there is also evidence for the cross-modal reorganization to take place when blindness develops late in life: although reduced with respect to those observed in early-blind patients, both auditory and tactile activations of the visual cortices have been described in late-blinds (Burton, 2003; Cunningham et al., 2011, 2015a,b; Voss et al., 2006), and one study showed that the extent and strength of tactile-evoked responses in V1 correlate with vision loss in late-blind individuals affected with retinitis pigmentosa (RP) (Cunningham et al., 2011, 2015b).

The underlying anatomical pathways supporting cross-modal responses may differ between early and late-blind individuals (Collignon et al., 2009). For example, Collingon et al. (2013) found that the auditory activity in the occipital cortex of congenitally blind individuals was most likely conveyed by direct connections from A1 to V1, whereas in late–blind individuals it appeared to be conveyed by feedback projections from multisensory parietal areas. The specific mechanisms underlying the rerouting of non-visual information are still unclear (Voss, 2019), however they might deeply differ depending on the age of blindness onset: while non-visual connections to the occipital cortex in congenital/early-blind individuals might be due to the lack of pruning typically triggered by visual experience, cross-modal responses in late-blind individuals might reflect the unmasking of non-visual pre-existing connections (normally supporting multisensory integration in sighted individuals) which can be progressively reinforced and result in permanent structural changes with new synapsis formation (Merabet et al., 2008).

Whatever the cause or the precise mechanisms underlying cross-modal plasticity might be, it can potentially interfere with the outcome of restoration techniques (Collignon et al., 2011a; Merabet et al., 2005). This possibility has been clearly demonstrated in deaf children who underwent cochlear implant: sustained and prolonged periods of deafness induced stronger cross-modal reorganization of acoustic cortex (as measured by glucose hypometabolism) and hampered recovery after cochlear implant (Lee et al., 2001). Although the predictive link between the extent of cross-modal reorganization and successful outcome of vision restoration still has to be directly demonstrated, one potential prognostic predictor of treatment effectiveness might be V1 cortical thickness: Aguirre et al (2016) showed that this parameter is strictly related to the strengths of cross-modal responses, independently of the patient age and blinding pathology.

Thus, cross-modal reorganization might pose a major challenge to vision restoration therapies, and if we take into account the lower plasticity in adulthood, it seem even more difficult to imagine successful visual recovery in late-blinds. On the other hand however, it is possible that sight restoration might be more feasible in late than in early-blind individuals, because visual cortex wiring successfully occurred before the disease in late–blind individuals. A recent study showed that functional connectivity between visual areas is still retinotopically organized in blind patients affected with juvenile macular degeneration, even after prolonged period of visual deprivation, directing further hope on the possibility to have positive outcomes from sight restoration techniques that can rely on a relatively intact visual system (Haak et al., 2016).

Yet, while these findings are encouraging, they do not guarantee that the adult visual brain retains the plastic potential necessary to mask the non-visual inputs potentiated or newly created during the blindness period and to boost the responses to the restored visual input; nor is it obvious that the visual system would 'learn to see again' with an artificial and monocular input.

### 3.3 Cortical plasticity after sight restoration

The first attempts to restore vision in adulthood focused on early-blind patients and showed limited visual recovery in these patients (Dormal et al., 2015; Fine et al., 2003; Gregory and Wallace, 1963; Saenz et al., 2008). Patient MM became blind at the age of three years old, and once his sight was restored in his 40ies, he was only able to recover visual abilities strictly related to the visual experience before blindness, such as perception of simple forms, color and motion, while perception of more complex 3D forms, objects and faces remained severely impaired even after several years of restored vision (Fine et al., 2003; Huber et al., 2015). fMRI studies on patient MM and on another early-blind patient whose vision was partially restored in adulthood, showed that cortical plasticity was also limited although not completely absent: several months after vision restoration, cross-modal auditory responses continued to coexist with the restored visual activations in area V1 (Dormal et al., 2015) and MT (Saenz et al., 2008). Interestingly however Dormal et al. (2015) observed that cross-modal responses in extra striate areas decreased after surgery and vision improved suggesting that cross-modal reorganization can be partially reversed in their early-blind patient.

Very few studies tracked the cortical reorganization process in late-blind patients after vision restoration. One study tracked the cortical responses in a 80 years old woman with wet age-related AMD undergoing intravitreal antiangiogenic injections (ranibizumab) over about 1 year period (Baseler et al., 2011a). Microperimetry showed that the scotoma decreased over time, visual acuity, fixation stability and reading skills improved as well. Interestingly after the first treatment, BOLD responses elicited by full-field flickering lights showed a tendency to be located in more posterior occipital regions, corresponding to the patient scotoma, however without filling it completely. Cunningham et al (2015a) tested two late-blind patients affected by retinitis pigmentosa (RP) implanted with Argus II Retinal Prosthesis and showed that the strength of tactile-evoked responses in V1 depended from the time from surgery: cross-modal activations were much reduced in the patient implanted for fifteen weeks before the scanning with respect to the patient who had the implant only for six weeks (Fig 2A). No visual responses were detected in this study, potentially due to the relatively short time period between the implantation and the scanning. A more recent study tested RP patients implanted with the same system at least six months after surgery and found that visual responses to flashes of lights increased in LGN and V1 after the surgery (Fig. 2B) (Castaldi et al., 2016). Importantly visual recovery (quantified as the behavioral performance on a challenging detection task) depended on the time from surgery and practice with the device (Fig. 2C) and was mirrored by enhanced BOLD responses to flashes after surgery, suggesting that the activity measured in visual areas had a functional relevance. Interestingly some weak visual activations were present already before the surgery, although participants never reported perceiving the flashing lights during the scanning. In particular, responses in extra striate areas were stronger before the surgery, whereas V1 showed stronger activations after the surgery. It is possible that even before surgery some spared visual input reached patients' V1, given the consistent albeit small BOLD response observed. However these visual signals might have been actively suppressed by extra striate areas, probably because the visual stimulation was not sufficiently reliable, aberrant and delayed. The suppression could have cross-sensory or motor origin, but also visual given that direct thalamic visual input to associative cortex have repeatedly observed in normal brain (Ajina et al., 2015; Tamietto and Morrone, 2016). Animal models reported spontaneous ganglion cells hyperactivity during photoreceptors degeneration process (Ivanova et al., 2016; Trenholm and Awatramani, 2015; Zeck, 2016). Suppressing such paroxysmal discharges might be beneficial to prompt faster cross-modal reorganization. This possibility is in line with the idea of a dysfunctional gating mechanism in blinds which would allow feedback projections to freely interact with the incoming input signal (Masuda et al., 2010), even when this is noisy and aberrant and the nature of this interaction might be inhibitory.

Taken together the results of these experiments suggest that when visual signals are restored in late-blinds, not only cross-modal responses in V1 needs to be attenuated or eliminated, but also the suppression of the incoming visual input to V1 should be released, and both these processes might take a long time. Yet, the decreased tactile-evoked responses (Cunningham et al., 2015a) and the increased visual BOLD signal in V1 (Castaldi et al., 2016) in patients that used the prosthesis for longer time, suggest that a spared plastic potential is retained by the adult visual brain and encourage continuative research to overcome the major obstacles limiting the expected outcome from vision restoration techniques. These obstacles include, among others, the limited quality of visual percepts that can be obtained (at best) with the current technology - for a review on advantages and limitations of different vision restoration methodologies see: Fine and Boynton (2015). For example, simulation of the likely perceived images with epiretinal devices showed that they might look extremely distorted because the electrode array stimulate unspecifically axons of the ganglion cells together with their cellular bodies (Fine and Boynton, 2015). It is thus not surprising that patients find particularly difficult to recover complex aspects of visual perception such as correctly discriminating the direction of drifting gratings (Castaldi et al., 2016). Although the technology can certainly be further improved, several studies have already shown that, after extensive training, RP patients implanted with retinal prosthesis can learn to perform some easy task, which can be nevertheless important for the patients' quality of life, such as moving independently in the space, locating large sources of light and even read large highly contrasted letters (Barry et al., 2012;

Castaldi et al., 2016; Chader et al., 2009; da Cruz et al., 2013; Dorn et al., 2013; Rizzo et al., 2014). Improvement of visual acuity and visual field perimetry were reported in patients implanted with both epiretinal and subretinal implants (Chow 2013; Chow et al., 2004, 2010; Rizzo et al., 2014, 2015). Interestingly visual field improvement extended outside the retinotopic regions directly stimulated by the implant (Rizzo et al., 2014) and even in the unoperated eye (Rizzo et al., 2015). The fact that the visual field recovering is not strictly limited to the stimulation site might suggest that the neuroplastic response acts 'peripherally' and originate from the release of retinal trophic factors induced by the current injection which diffuse to non-stimulated regions of the retina (Ciavatta et al., 2009). However it is also possible that the artificial vision provided by retinal implants reopens visual plasticity at 'central level', which would better explain the visual recovery of the fellow eye observed at least in one study (Rizzo et al., 2015). Perhaps the fellow eye benefits from an anterograde effect mediated by cross talk of neural discharges along the optic nerve after the chiasma or at the level of LGN where the projections from the two eyes are closely interlayered. Interestingly some increase of BOLD response have also been observed at LGN level after prolonged use of the prosthetic devise (Castaldi et al., 2016).

In sum, despite long time and extensive training is needed to recover a functional use of artificial vision and although the specific source of such neuroplastic responses is currently unknown, the reviewed results suggest that vision restoration techniques can rely on residual neuroplasticity retained by the adult brain and that, especially for late-blinds, it might be possible to restore vision even after several years of blindness.

## 4. Conclusions

We reviewed behavioral and neural evidence suggesting that a considerable degree of neuroplasticity is preserved well beyond the closure of the critical periods for vision. Behavioral and neural evidence indicates that in sighted individual short periods of monocular deprivation can trigger homeostatic plasticity and that this strategy can be successfully used to improve visual perception in amblyopic adults. Evidence for cross-modal reorganization of the visual brain of blind individuals is compelling and recent findings suggest that this process can be reverted even after several years of blindness. However it is important to acknowledge the fact that this neuroplasticity is a very slow process and that the quality of the regained visual perception is limited. Yet, the field of visual restoration techniques is still at the beginning and much improvement can be expected in the upcoming years. Certainly neuroscientists can contribute to the development of this field by exploring the properties, characteristics and mechanisms of the spared plastic potential retained by the human visual brain to calibrate optimally vision restoration therapies and maximize their expected outcome.

Perhaps, the most outstanding question to answer is whether we can boost neuroplasticity to make visual recovery more effective and rapid. Although the answer might depend on the age of blindness onset and on the cause determining the visual deficit, identifying some common principles and mechanisms guiding neuroplasticity during development and adulthood and across modalities mighty lead to develop strategies effective under a wide range of circumstances and diseases. Hopefully in the next decades we will have the complete approach to rescue visual function, facilitating the adult brain to learn to "see" again, even after several years of blindness or abnormal visual experience.

## Acknowledgments

This research has received funding from Fondazione Roma under the Grants for Biomedical Research: Retinitis Pigmentosa (RP)-Call for proposals 2013 (http://www.fondazioneroma.it/it/index.html, http://wf-fondazioneroma.cbim.it/), project title: Cortical Plasticity in Retinitis Pigmentosa: an Integrated Study from Animal Models to Humans, from the European Research Council under the European Union's Seventh Framework Programme (FPT/2007-2013) under grant agreement No. 338866 ECPLAIN (http://www.pisavisionlab.org/index.php/ projects/ecsplain) and under the European Union's Horizon 2020 research and innovation programme (grant agreement No 801715 – PUPILTRAITS).

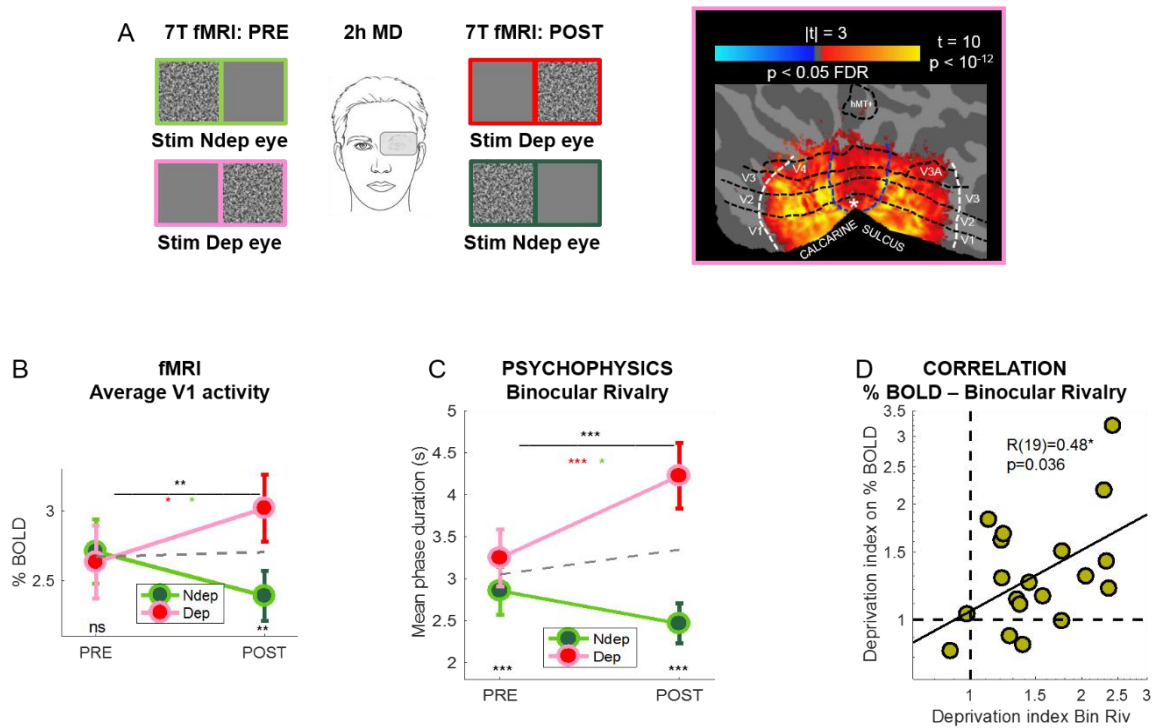

Figure 1. Short-term visual deprivation induces functional reorganization of cortical circuits in sighted adult humans.

(A) During 7T fMRI scanning participants' eyes were separately stimulated with band-pass noise and BOLD responses were measured before and after 2h of monocular deprivation. The flat map shows extensive BOLD response before deprivation elicited by the noise stimuli in all visual areas. (B) The percentage BOLD signal measured in V1 elicited by stimulation of the deprived eye and non-deprived eye was comparable before deprivation, while respectively enhanced and reduced after 2h of monocular deprivation. (C) The neural changes described are reflected at behavioral level: mean phase duration during binocular rivalry are balanced across the two eyes before deprivation, while mean phases are longer/shorter after deprivation for the deprived/non-deprived eye respectively. (D) These perceptual effects correlate with the deprivation index measured with fMRI. (A-D) Reproduced from Figure 1 and Figure 3 (Binda et al., 2018), eLife, published under the Creative Commons Attribution 4.0 International Public License (CC BY 4.0; https://creativecommons.org/licenses/by/4.0/)."

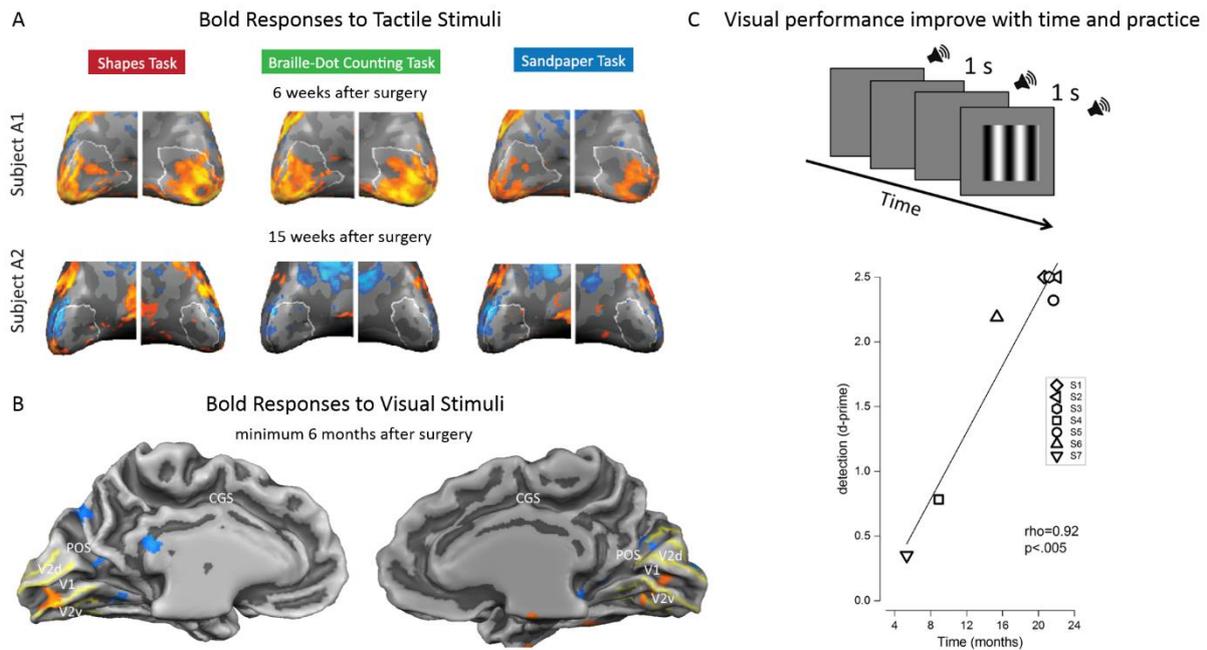

Figure 2. After sufficient time and practice the adult visual brain can learn to see again with artificial vision.

(A) BOLD responses elicited by three different tactile tasks in two RP patients implanted with Argus II retinal prosthesis (subjects A1 and A2) who underwent fMRI scanning after 6 and 15 weeks after surgery. Participants were required to haptically determine the symmetry of raised-line shapes (shape task), count the number of dots in Braille letters (Braille-dot counting task), and evaluate the roughness of sandpaper discs (sandpaper task). Independently of the task, strong tactile evoked responses can be observed in occipital cortex after only 6 weeks after surgery (subject A1). The occipital cross-modal activation is much reduced after 15 weeks from surgery (subject A2). (B) BOLD responses elicited by visual stimuli (flashes of lights) before (blue) vs after (red) surgery in a group of RP patients implanted with Argus II. After the surgery visual BOLD signal is enhanced in V1. Crucially participants were scanned at least 6 months after the surgery. (C) RP patients were asked to choose the interval, demarcated by sound, in which a large high contrast grating was presented. The behavioral performance in a contrast detection task improves as a function of time from surgery. Patients needed time and practice to learn how to interpret and use the restored visual input. (A) Reproduced with permission from Figure 3 from Cunningham SI, Shi Y, Weiland JD, et al. Feasibility of structural and functional MRI acquisition with unpowered implants in Argus II retinal prosthesis patients: a case study. Trans Vis Sci Tech. 2015;4(6):6., ARVO copyright holder. (B, C) Reproduced from Figure 2 and Figure 5 from Castaldi et al. (2016) Visual BOLD Response in Late Blind Subjects with Argus II Retinal Prosthesis. PLOS Biology 14(10): e1002569. https://doi.org/10.1371/journal.pbio.1002569, published under the Creative Commons Attribution license (CC BY).